\documentclass[twocolumn,prb,amsmath,amssymb]{revtex4}

\usepackage{graphics}
\usepackage{graphicx}
\usepackage{dcolumn}
\usepackage{bm}
\usepackage{longtable}

\begin{document}

\title{Translocation energy of ions in nano-channels of cell membranes}

\author{Sofian Teber}

\email{steber@ictp.trieste.it}

\affiliation{The Abdus Salam ICTP, Strada Costiera 11, 34014, Trieste, Italy}

\date{\today}

\begin{abstract}
Translocation properties of ionic channels are investigated, on the basis of classical electrostatics, with an emphasis on asymptotic formulas for the potential and field associated with a point charge in the channel. Due to image charges in the membrane, we show that ions in an infinite length channel interact via a one-dimensional (1D) Coulomb potential. The corresponding  electrostatic barrier $\Sigma$ is characterized by a ``geometric mean"  screening $\Sigma \propto e^2 / \sqrt{\epsilon_w \epsilon_m}R$ ($R$ being
the radius of the pore, and $\epsilon_m \approx 2$ and $\epsilon_w
\approx 80$ the room temperature dielectric constants of membrane
and water, respectively). There exists a crossover length, $x_0
\propto R \sqrt{\epsilon_w / \epsilon_m} \sim 6.3 R$, below which
the 1D potential governs the electrostatics and beyond
which the three-dimensional (3D) Coulomb potential screened by the
membrane takes over. Knowledge of this length enables us to
discriminate between long channels, the length $L$ of which
satisfies: $L \gg 2 x_0$, and short channels for which $L \ll 2
x_0$. The latter condition is satisfied by most realistic channels
({\it e.g.}, gramicidin A where $R \approx 3 {\mathrm{\AA}}$, $L
\approx 2.5 {\mathrm{nm}}$ and $2x_0 \approx 3.8 {\mathrm{nm}}$)
whose translocation energy is therefore controlled by the part of
the self-energy, $\Sigma$, arising from the 1D potential. On
this basis, we derive an expression for $\Sigma$, with no fitting parameter, which applies to a generic nano-channel of length $L$ and radius $R$. Our results are related to model-independent translocation properties of nano-scale ionic channels, they improve on previous, curve-fitting, formulas and agree to within $5\%$ with estimates, resulting from numerical
simulations, available in the literature on the subject.
\end{abstract}

\maketitle


\section{Introduction}

\begin{figure}
\includegraphics[width=8cm,height=4cm]{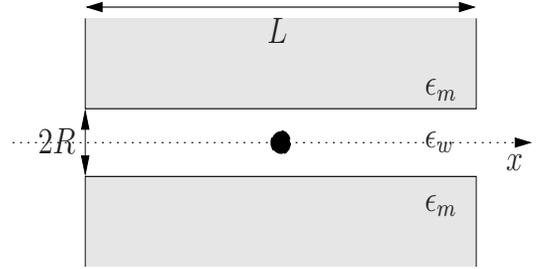}
\caption{ \label{Channel} Schematic view of a channel of length
$L$, radius $R$ with an ion sitting in its center, {\it i.e.} at $x=0$. The channel
consists of water, of dielectric constant $\epsilon_w \approx 80$,
and is surrounded by the membrane, of dielectric constant
$\epsilon_m \approx 2$. A typical channel we are interested in, {\it e.g.} gramicidin A, has a length $L \approx 2.5 {\mathrm{nm}}$ and a radius $R \approx 3 {\mathrm{\AA}}$.}
\end{figure}

The simplest physical models of bilayer membranes turn out to be
particularly useful to understand the basic electrostatic interplays
between the membrane and the ionic flow.  Consider the phospholipid bilayer
as a thin slab of non-conducting material separating two aqueous solutions, and
an aqueous ion channel as a cylindrical region connecting the two solutions, see
Fig~\ref{Channel}. One apparent obstacle to ionic conduction is
the electrostatic energy barrier which has to be passed in order
to move the ion through the channel. In 1969, Parsegian
\cite{Parsegian} found that, for an infinite channel, the
electrostatic self-energy $\Sigma$, {\it i.e.} the energy
required to move an ion from infinity to the center of the
channel, is of the order of $16~{k_BT}$. A major
contribution to this self-energy comes from the self-image of the
charge due to the consequent difference between the dielectric
constant of the water channel ($\epsilon_w \approx 80$, at room
temperature) and that of the lipid membrane ($\epsilon_m \approx
2$). Such a large barrier would prevent {\it any} ionic flow, contrary
to the relatively high conductances observed in experiments
\cite{IC_book}. Moreover, these are compatible with barriers not
exceeding $6~{k_BT}$. The most obvious limitation of previous models
is that they consider channels of infinite length (no screening by counter
ions is assumed because typical channel radii considered here are of the order of
a few ${\mathrm{\AA}}$). Taking into account the finite length of
the channel (gramicidin A has a characteristic length of
$25~{\mathrm{\AA}}$), as well as a high dielectric shield of the
channel, the authors of Refs. \onlinecite{Levitt}
and~\onlinecite{Jordan} have shown, with the help of numerical
methods, a considerable decrease of this barrier, reaching
$6.7~{k_BT}$.  A large number of recent studies also
address this question, generally focusing on specific models and
taking into account of the channel nano-structure, using numerical
simulations of the Poisson-Boltzmann or Poisson-Nernst-Planck
type, to improve on previous estimates~\cite{IC_book}.

In this contribution, the original, model-independent, problem of
the electrostatics of point-like ions in an infinite channel is
reconsidered, with an emphasis on asymptotic expressions for the
potential and field which have not been derived previously. Our
quantitative results show the appearance of a length scale: $x_0
\propto R \sqrt{\epsilon_w / \epsilon_m}$, which separates
regions, in space, where the effect of the image-charges on the
ion at the origin are crucially different. At distances $x$, along
the channel, smaller than $x_0$ ($x \ll x_0$), the electrostatic
potential, $\varphi$, is one-dimensional (1D) and displays an unusual, {\it
i.e.} algebraic in the dielectric constants, screening: $\varphi
\propto (e^2 / \sqrt{\epsilon_w \epsilon_m} R) (1 - |x|/x_0)$. At
distances $x$, along the channel, larger than $x_0$ ($x \gg x_0$),
the electrostatic potential is three-dimensional (3D) and everywhere, even in the
water channel, screened by the membrane: $\varphi \propto e^2 /
\epsilon_m |x|$. In the region $x \ll x_0$, the ions are
anti-confined, {\it i.e.} subject to a constant repulsive force.
The crossover length, $x_0$, is found to be of the order of, or
larger than, the average length of a typical ionic channel and our
asymptotic expressions, in the 1D Coulomb regime, thus
capture the essence of the phenomena characteristic of realistic,
{\it i.e.} nano-length, channels. On this physical basis, we will
therefore derive a formula for $\Sigma$ depending on the length
$L$ and radius $R$ of a typical pore. This formula for
$\Sigma$, will be shown to agree, within $5\%$ of accuracy, with
the available numerical results in the field and therefore
improves on previous, curve-fitting, expressions. This
will lead us to an accurate knowledge, of the main exponential
dependence, of the conductance $G$, of finite-length channels: $G
\propto \exp[-\Sigma / k_B T]$.

This article is organized as follows. We first consider the case
of an infinite channel where calculations may be carried out
straightforwardly. In Sec.~\ref{potential_in}, the electrostatic
potential, within the infinite channel, is derived in various
asymptotic regimes. The length $x_0$ is defined. In
Sec.~\ref{potential_out}, similar asymptotic formulas for the
potential in the membrane are derived. These results are used, in
Sec.~\ref{field}, to derive the components of the electric field. In
Sec.~\ref{self_energy}, the image self-energy of a charge in an
infinite channel is derived and the part of this self-energy,
arising from the 1D Coulomb potential, is singled out. In
Sec.~\ref{estimates} numerical estimates of the various
quantities, derived in the previous sections, are given and
compared with exact numerical simulations. In
Sec.~\ref{finite_sizes}, the case of finite size channels is
considered and, on the basis of the results of previous sections,
an expression for the self-energy of nano-channels is derived. A
comparison is then made with the available results in the
literature, based on numerical simulations. Finally, the
conclusion is given in Sec.~\ref{conclusion}.


\section{Electrostatic potential inside an infinite channel}
\label{potential_in}

\begin{figure}
\includegraphics[width=8cm,height=4cm]{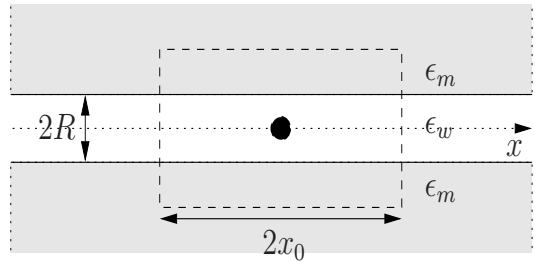}
\caption{ \label{InfC} Schematic view of an infinite channel,
including the cylinder of length $2x_0$ and radius $x_0$, cf. Eq.~(\ref{x_0}).}
\end{figure}

Consider a point-like ion (charge $e$) in an infinite cylinder of
radius $R$ and of dielectric constant $\epsilon_w$ (of water),
embedded in a media (the membrane) of dielectric constant
$\epsilon_m \ll \epsilon_w$, cf. Fig.~\ref{InfC}.
The electrostatic potential, $\varphi(x,{\bf r_\bot})$, of such
an ion {\it within} the cylinder is given by:
\begin{eqnarray}
\label{phi_real}
\varphi(x,{\bf r_\bot}) = \int_{0}^{+\infty}{{dk \over \pi}} \varphi(k,{\bf r_\bot}) \cos(kx),
\end{eqnarray}
where $k$ is the wavenumber along the (x-)axis of the channel and:
\begin{eqnarray}
\label{phi_fourier}
&&\varphi(k,{\bf r_\bot}) = {2 e \over \epsilon_w} K_0(kr_\bot) +
\nonumber \\
&&{2 e \over \epsilon_w} {(\epsilon_w -
\epsilon_m)K_0(kR)K_1(kR)I_0(kr_\bot) \over \epsilon_w
K_0(kR)I_1(kR) + \epsilon_m K_1(kR)I_0(kR)},
\end{eqnarray}
where $I_{\nu}$ and $K_{\nu}$ are modified Bessel functions of the
first and second kind, respectively. The result of
Eqs.~(\ref{phi_real}) and~(\ref{phi_fourier}) may be
straightforwardly derived by solving the appropriate inhomogeneous
Poisson equation. The first term in Eq.~(\ref{phi_fourier})
corresponds to the usual ($1/r$ in 3D) Coulomb potential of a
charge in an infinite media of dielectric constant $\epsilon_w$
and the second term corresponds to the potential due to the image
charges (repulsive, as it should be, for $\epsilon_w >
\epsilon_m$, because the charge is sitting in the high dielectric
constant media). A similar expression has been derived
elsewhere~\cite{Sm} and we shall therefore not prove it here.

At distances smaller than the radius of the channel, $kR \gg 1$, we may
use the asymptotic expressions of modified Bessel functions~\cite{GR}:
\begin{eqnarray}
\label{as_bessel}
K_\nu(kR) \approx {\exp(-kR) \over \sqrt{2 \pi k R}}, \quad
I_\nu(kR) \approx {\exp(kR) \over \sqrt{2 k R / \pi}}.
\end{eqnarray}
Substituting these expressions in the second term, the image part, of
Eq.~(\ref{phi_fourier}), leads to a term $\approx e/ \epsilon_w R$, which,
at $x \ll R$, is negligible compared to the first one: $e/ \epsilon_w |x|$.
At distances, $x \ll R$, the images play no role, and the Coulomb potential
is the usual 3D one, screened by the dielectric constant of water:
\begin{eqnarray}
\label{pot_xsmR}
\varphi({\bf r}) = {e \over \epsilon_w r}, \quad r \ll R.
\end{eqnarray}
Because all phenomena we shall be interested in, in the following, are related
to the effect of image charges, $R$ will be our lowest distance cut-off. We
will therefore explore the properties of the system at distances larger than
the radius of the channel.

\begin{figure}
\includegraphics[width=8cm,height=4cm]{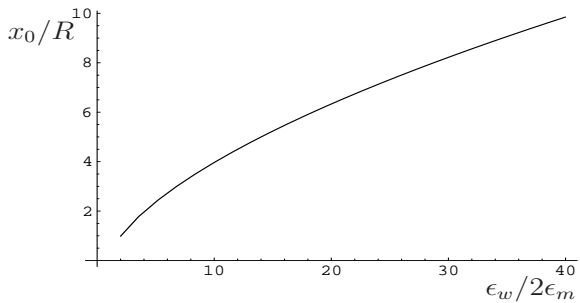}
\caption{ \label{lengthx0} The crossover length $x_0$, in units of the radius of the
channel $R$, as a function of the ratio $\epsilon_w / 2 \epsilon_m$.}
\end{figure}

\begin{figure}
\includegraphics[width=8cm,height=4cm]{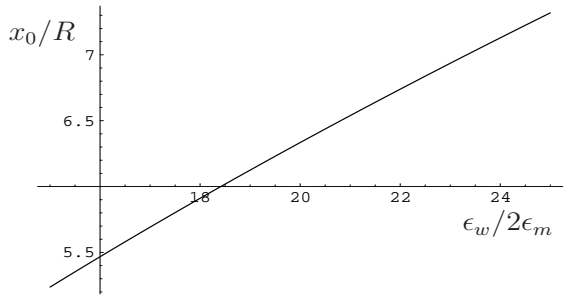}
\caption{ \label{lengthx02} The crossover length $x_0$, in units of the radius of the
channel $R$, as a function of the ratio $\epsilon_w / 2 \epsilon_m$ in the vicinity of the
realistic value $\epsilon_w / 2 \epsilon_m = 20$, for a water
channel in a lipid membrane ($\epsilon_w \approx 80$ and
$\epsilon_m \approx 2$).}
\end{figure}

We focus on the asymptotics of Eq.~(\ref{phi_fourier}) at
distances, along the channel, larger than its radius, {\it i.e.}
$kR \ll 1$. We will need, in the following, the series expansions of
Bessel functions, which may be found in any standard table, {\it e.g.} Ref.
\onlinecite{GR}. To the highest order of main interest to us, these
series expansions read:
\begin{eqnarray}
\label{series_bessel}
&&K_0(kR) \approx \log(2/kR) - \gamma, \nonumber \\
&&K_1(kR) \approx -(\log(2/kR) - \gamma) kR / 2, \nonumber \\
&&I_0(kR) \approx 1 + (kR / 2)^2 + ..., \nonumber \\
&&I_1(kR) \approx kR/2 + ...,
\end{eqnarray}
where $\gamma \approx 0.577$ is Euler's constant. With the help of Eq.~(\ref{series_bessel}),
Eq.~(\ref{phi_fourier}) may then be approximated by:
\begin{eqnarray}
\label{phi_real_series}
&&\varphi(x,{\bf r_\bot}) \approx {e \over \epsilon_w |x|} +
\nonumber \\
&&{2e \over \epsilon_m R}(1-{\epsilon_m \over \epsilon_w})
\int_{0}^{+\infty}{{d\xi \over \pi}} {\xi K_0(\xi)K_1(\xi) \cos(\xi x/R)
\over 1 + (\xi x_0 /R)^2},
\end{eqnarray}
where the dimensionless $\xi = kR$ has been introduced and a
characteristic distance, $x_0$, appears and reads:
\begin{eqnarray}
\label{x_0}
x_0 = R \sqrt{{\epsilon_w \over 2 \epsilon_m}} \sqrt{\log(2x_0/R) - \gamma}.
\end{eqnarray}
For clarity, the solutions of Eq.~(\ref{x_0}) for the dimensionless ratio $x_0/R$ as a function of
$\epsilon_w / 2 \epsilon_m$ are plotted on Fig.~\ref{lengthx0}. Fig.~\ref{lengthx02} provides more
accurate values for the crossover length $x_0$ in the vicinity of the realistic value $\epsilon_w / 2 \epsilon_m = 20$, for a water channel in a lipid membrane ($\epsilon_w \approx 80$ and
$\epsilon_m \approx 2$).

In the limit $\xi \ll R/x_0$, {\it i.e.} $x \gg x_0$, Eq.~(\ref{phi_real_series}) reduces to:
\begin{eqnarray}
\label{phi_real_series_largex}
&&\varphi(x,{\bf r_\bot = 0}) \approx {e \over \epsilon_w |x|} +
\nonumber \\
&&{2e \over \epsilon_m R}(1-{\epsilon_m \over \epsilon_w})
\int_{0}^{+\infty}{{d\xi \over \pi}} \xi K_0(\xi)K_1(\xi) \cos(\xi x/R).
\end{eqnarray}
The integral is a standard one and, as a result:
\begin{eqnarray}
\label{phi_real_series_largex2}
&&\varphi(x,{\bf r_\bot}) \approx {e \over \epsilon_w |x|} +
{e \over \epsilon_m |x|}(1-{\epsilon_m \over \epsilon_w}) = {e \over \epsilon_m |x|},
\nonumber \\
&&x \gg x_0, \quad r_\bot < R,
\end{eqnarray}
which corresponds to the usual 3D Coulomb interaction screened by the membrane.

\begin{figure}
\includegraphics[width=8cm,height=5cm]{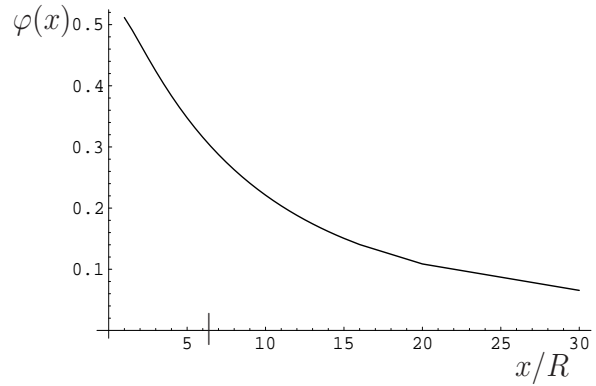}
\caption{ \label{Pot} The electrostatic potential at the center of an infinite channel ($r_\bot = 0$) as a function of $x/R$, the dimensionless distance along the water channel ($\epsilon_w \approx 80$ and $\epsilon_m \approx 2$). The tick at $x_0 = 6.3 R$ marks the crossover between the 1D-potential, $x \ll x_0$, and the 3D potential, $x \gg x_0$.}
\end{figure}

In the limit $\xi \gg R/x_0$, {\it i.e.} $R \ll x \ll x_0$, Eq.~(\ref{phi_real_series}) reads:
\begin{eqnarray}
\label{phi_real_series_smallx2}
&&\varphi(x,{\bf r_\bot}) =
{e \over \epsilon_m x_0} K_0(R/x_0)J_0(r_\bot/x_0)\exp(-|x|/x_0),
\nonumber \\
&&R \ll x \ll x_0, \quad r_\bot < R,~~
\end{eqnarray}
where corrections in $x/x_0$, $r_\bot/x_0$ as well as $R/x_0$ have
been summed, to all orders, into the special functions appearing
in  Eq.~(\ref{phi_real_series_smallx2}). With the help of series
expansions of these special functions, to the lowest orders in the
small quantities above: $J_0(r_\bot/x_0) \approx 1$ and:
$K_0(r_\bot/x_0) \approx -\gamma + \log(2 x_0/R)$,
Eq.~(\ref{phi_real_series_smallx2}) yields:
\begin{eqnarray}
\label{phi_real_series_smallx}
&&\varphi(x,{\bf r_\bot}) \approx
{e \over  \epsilon_m x_0} \left( \log(2x_0/R) - \gamma \right)
\left(1 - {|x| \over x_0} \right),
\nonumber \\
&&R \ll x \ll x_0, \quad r_\bot < R.
\end{eqnarray}
Eq.~(\ref{phi_real_series_smallx}) shows that the potential is 1D-like, because of the linear dependence on the coordinate, $x$, along the channel. The electrostatic potential along the axis of the channel has been plotted on Fig.~\ref{Pot}, from Eqs.~(\ref{phi_real}) and (\ref{phi_fourier}).

%

\section{Electrostatic potential outside the infinite channel}
\label{potential_out}

In this section, we will still focus on the ion {\it within} the
pore but determine the potential due to this ion {\it outside} the
pore, {\it i.e.} for $r_\bot > R$. This potential reads:
\begin{eqnarray}
\label{phi_fourier_out} \varphi(k,{\bf r_\bot}) = {2 e \over
kR} {K_0(kr_\bot) \over \epsilon_w K_0(kR)I_1(kR) + \epsilon_m
K_1(kR)I_0(kR)}.
\end{eqnarray}
Being interested in distances, along the axis of the pore, which
are larger than $R$, {\it i.e.} $kR \ll 1$,
Eq.~(\ref{phi_fourier_out}) reads, in this limit:
\begin{eqnarray}
\label{phi_out_series}
\varphi(x,{\bf r_\bot}) = {2 e \over \epsilon_m x_0}
\int_{0}^{+\infty}{{d\xi \over \pi}} {K_0(\xi r_\bot / x_0) \cos(\xi x/x_0) \over 1 + \xi^2},
\end{eqnarray}
where the dimensionless variable $\xi = kx_0$ and $x_0$ is given by Eq.~(\ref{x_0}).

 For $x \gg x_0$, we recover the potential of a 3D point-like charge:
\begin{eqnarray}
\label{phi_out_largex}
\varphi(x,{\bf r_\bot}) = { e \over \epsilon_m r}, \qquad x \gg x_0, \quad \forall~r_\bot > R,
\end{eqnarray}
where $r=\sqrt{x^2+r_\bot^2}$ and the screening is by the dielectric constant of the membrane.

 For $x \ll x_0$, the potential reads:
\begin{eqnarray}
\label{phi_out_smallx}
&&\varphi(x,{\bf r_\bot}) = { e \over \epsilon_m x_0} {\pi \over 2} \left[ H_0(r_\bot / x_0)
- Y_0(r_\bot/x_0) \right],
\nonumber \\
&&x \ll x_0, \quad r_\bot > R,
\end{eqnarray}
where $H_0$ is a Struve function and $Y_0$ is a Neumann function.
Using asymptotic and series representations of these special
functions~\cite{GR}, we find that:
\begin{eqnarray}
\label{phi_out_smallx_smallr}
&&\varphi(x,{\bf r_\bot}) \approx { e \over \epsilon_m x_0} \left[ \log(2x_0/r_\bot)-\gamma \right],
\nonumber \\
&&x \ll x_0, \quad  R < r_\bot \ll x_0,
\end{eqnarray}
and that:
\begin{eqnarray}
\label{phi_out_smallx_larger}
\varphi(x,{\bf r_\bot}) = { e \over \epsilon_m r_\bot},
\qquad x \ll x_0, \quad  r_\bot \gg x_0.
\end{eqnarray}
Eq.~(\ref{phi_out_smallx_smallr}) shows that, inside the shell of
radius $x_0$ outside the pore, the potential is 2D-like. As for
the 1D potential of Eq.~(\ref{phi_real_series_smallx}), this
implies that field lines are mostly located within the water
channel because of its high dielectric constant. Beyond $x_0$,
such a potential crosses over to the 3D Coulomb potential of
Eq.~(\ref{phi_out_smallx_larger}), similarly to what happens
inside the channel, cf. Eq.~(\ref{phi_real_series_largex2}).

The potential outside the pore of
Eq.~(\ref{phi_out_smallx_smallr}) has also an $x-$dependence which
we have neglected as a first approximation. Taking this dependence
into account, the latter may formally be expressed as:
\begin{eqnarray}
\label{phi_out_smallx_smallr2}
&&\varphi(x,{\bf r_\bot}) = { e \over \epsilon_m x_0} K_0(r_\bot/x_0) \exp(-|x|/x_0),
\nonumber \\
&&x \ll x_0, \quad  R < r_\bot \ll x_0.
\end{eqnarray}
The series expansions of the special functions appearing in
Eq.~(\ref{phi_out_smallx_smallr2}), give the dominant
contributions in the small parameters $x/x_0$ and $r_\bot/x_0$,
{\it e.g.} the exponential gives the linear dependence on $|x|$
and the MacDonald function $K_0$ gives the logarithmic dependence
on $r_\bot$.


\section{Electric field in an infinite channel}
\label{field}

The previous sections have shown the importance of the crossover
length, $x_0$, which separates regions, in space, characterized by
different sectors of the electrostatic potential
controlled by the effect of image charges. We proceed now on
deriving the corresponding components of the electric field of the ions in the channel.

Within the channel and at distances, from the ion at the origin,
smaller than $x_0$, cf. Fig.~\ref{InfC}, the electrostatic
potential is given by Eq.~(\ref{phi_real_series_smallx2}), which
corresponds, in the first order in $x/x_0$ to a 1D Coulomb
potential, cf. Eq.~(\ref{phi_real_series_smallx}). The
corresponding components of the electric field then formally
(expansions of the various functions involved have to be taken)
read:
%
%
\begin{eqnarray}
\label{electric_x_in}
&&E_x^{in} = {e \over \epsilon_m x_0^2} K_0(R/x_0)J_0(r_\bot/x_0)\exp(-|x|/x_0),
\nonumber \\
&&R \ll x \ll x_0, \quad r_\bot < R,
\end{eqnarray}
\begin{eqnarray}
\label{electric_perp_in}
&&E_\bot^{in} = {e \over \epsilon_m x_0^2} K_0(R/x_0)J_1(r_\bot/x_0)\exp(-|x|/x_0),
\nonumber \\
&&R \ll x \ll x_0, \quad r_\bot < R.
\end{eqnarray}
%
%
These expressions display the role of the image force: $J_0$ is
unity at the origin, so the longitudinal field is constant,
non-zero and positive. On the other hand, $J_1$ vanishes at the
origin, so there is no component of transverse force on the
charge. The 1D Coulomb force, which strongly prevents more
than one ion to be inside the pore, therefore reads:
\begin{eqnarray}
\label{conf_force}
F = e E_x^{in}(x=r_\bot =0) = {e^2 \over \epsilon_m x_0^2} (\log(2x_0/R) -
\gamma).
\end{eqnarray}
Another particular sector, of the electrostatic potential,
corresponds to the cylindrical shell surrounding the pore, at $R
\ll r_\bot \ll x_0$, still within the distance $x_0$ along the
channel, from the origin. In this region,
Eq.~(\ref{phi_out_smallx_smallr}) shows that the potential is
2D-like. With the help of Eq.~(\ref{phi_out_smallx_smallr2}),
which includes higher order corrections, the corresponding
components of the electric field then formally read:
%
%
\begin{eqnarray}
\label{electric_x_out}
&&E_x^{out} = {e \over \epsilon_m x_0^2} K_0(r_\bot/x_0) \exp(-|x|/x_0),
\nonumber \\
&&R \ll x \ll x_0, \quad  R < r_\bot \ll x_0,
\end{eqnarray}
\begin{eqnarray}
\label{electric_perp_out}
&&E_\bot^{out} = {e \over \epsilon_m x_0^2} K_1(r_\bot/x_0) \exp(-|x|/x_0),
\nonumber \\
&&R \ll x \ll x_0, \quad  R < r_\bot \ll x_0.
\end{eqnarray}
%
%
Finally, the last sector corresponds to the region outside the
cylinder of radius $x_0$ and length $2 x_0$ where the potential is
3D-like and isotropic (cf. Eqs.~(\ref{phi_real_series_largex2}),
(\ref{phi_out_largex}) and (\ref{phi_out_smallx_larger})). This
potential is still unusual, within the water channel, because, due
to the effect of image charges, it is screened by the membrane
only: $\varphi({\bf r}) = { e / \epsilon_m r},~(~ x >
x_0,~\forall~r_\bot~),~(~r_\bot > x_0,~\forall~x~)$. The electric
field then corresponds to the usual 3D electric field screened by
the dielectric constant of the membrane. We denote this field by:
\begin{eqnarray}
\label{electric_ext}
&&E^{ext} = { e \over \epsilon_m r^2} \nonumber \\
&&(~ x > x_0,~\forall~r_\bot~),~(~r_\bot > x_0,~\forall~x~).
\end{eqnarray}
%


\section{Image self-energy of an ion in an infinite channel}
\label{self_energy}

The total self-energy of a charge in the channel may be derived,
with the help of the self-potential, by the identity:
\begin{eqnarray}
\label{self_energy_def}
\Sigma^{(\infty)}= {e \over 2}  \varphi(0).
\end{eqnarray}
where the upper script $({\infty})$, refers to the fact that an
infinite channel is considered. As for the electrostatic potential,
in Eq.~(\ref{self_energy_def}), we consider the self-energy
arising due to image charges, the {\it image self-energy},
and measure it with respect to the Born energy of the charge
$e^2 / 2 \epsilon_w a$, where $a$ is the radius of the charge.
As for the electrostatic potential, we
decompose this image self-energy, by considering a part,
$\Sigma_{1}$, arising from the region $x \ll x_0$, and a part
related to larger distances, $\Sigma_{3}^{(\infty)}$, where the
potential is the regular 3D-like (but screened everywhere by the
membrane). Hence:
\begin{eqnarray}
\label{self_energy_inout}
\Sigma^{(\infty)} = \Sigma_{1}^{(\infty)} + \Sigma_{3}^{(\infty)}.
\end{eqnarray}

We focus first on the part arising from the 1D Coulomb, $\Sigma_{1}^{(\infty)}$. Suppose that the charge is brought to the origin along the channel. The corresponding self-potential is given by
Eq.~(\ref{phi_real_series_smallx2}) at $x=0$ and $r_\bot=0$. The
corresponding self-energy then reads:
\begin{eqnarray}
\label{self_energy_along_pore}
\Sigma_{1}^{(\infty)} \approx {1 \over 2} {e^2 \over \epsilon_m x_0} \left( \log(2x_0/R) - \gamma \right).
\end{eqnarray}
This self-energy may be evaluated by a similar way by
considering that the charge has been brought to the origin from
the membrane ({\it i.e.} perpendicular to the pore). Then,
Eq.~(\ref{phi_out_smallx_smallr2}) has to be used for the
self-potential which brings us back to
Eq.~(\ref{self_energy_along_pore}).

In order to determine the total image self-energy, and compare it
with exact numerical results, we need to take care of boundary
terms, arising from the cylindrical geometry of the system and
which influence the final value of the self-energy. We find it
therefore convenient to re-express Eq.~(\ref{self_energy_def}) in
the form:
\begin{eqnarray}
\label{self_energy_definition2}
\Sigma^{(\infty)} = {1 \over 2} e
\int{d{\mathcal V}} \rho({\bf r}) \varphi({\bf r}), \quad
\rho({\bf r}) = \delta({\bf r}),
\end{eqnarray}
where ${\mathcal V}$ is the total volume of the system. Using the
inhomogeneous Poisson equation, we may express the density of test
charge as a function of the potential, in
Eq.~(\ref{self_energy_definition2}), to obtain:
\begin{eqnarray}
\label{self_energy_definition2_3}
\Sigma = {1 \over 8 \pi} \int{d{\mathcal V}} \epsilon({\bf r}) ({\bf \nabla} \varphi({\bf r}))^2.
\end{eqnarray}
We focus first on contributions to the image self-energy, within $x_0$:
\begin{eqnarray}
\label{self_energy_decomposed}
\Sigma_{1}^{(\infty)} = { \epsilon_w  \over 8 \pi} \int_{in}{d{\mathcal V}} ({\bf E^{in}})^2 +
{ \epsilon_m  \over 8 \pi} \int_{out}{d{\mathcal V}} ({\bf E^{out}})^2.
\end{eqnarray}
The first term in Eq.~(\ref{self_energy_decomposed}), is
associated with the electric field inside the cylinder, cf.
Eqs.~(\ref{electric_x_in}) and (\ref{electric_perp_in}), and
reads:
\begin{eqnarray}
\label{self_energy_in} \Sigma_{1,in}^{(\infty)} = { \epsilon_w
e^2 \over 4 \epsilon_m^2 x_0} K_0^2({R \over x_0}) \int_{0}^{R
\over x_0}{d\xi_\bot} \xi_\bot \left[ J_0^2(\xi_\bot) +
J_1^2(\xi_\bot) \right],
\end{eqnarray}
where $\xi_\bot = r_\bot/x_0$. Expanding the integral in $R/x_0$ leads, in the lowest order in $R/x_0$, to:
\begin{eqnarray}
\label{self_energy_inin}
\Sigma_{1,in}^{(\infty)} =
{1 \over 2}{\epsilon_w e^2 R^2 \over 4 \epsilon_m^2 x_0^3} K_0^2(R/x_0),
\end{eqnarray}
where higher order terms in $R/x_0$ are neglected. Finally, using
Eq.~(\ref{x_0}), yields:
\begin{eqnarray}
\label{self_energy_in2}
\Sigma_{1,in}^{(\infty)} = {1 \over 4}{e^2 \over \epsilon_m x_0} \left( \log(2x_0/R) - \gamma \right),
\end{eqnarray}
With the same notations, the second term, in
Eq.~(\ref{self_energy_decomposed}), originates from the electric
field within $x_0$ but outside the cylinder, cf.
Eqs.~(\ref{electric_x_out}) and (\ref{electric_perp_out}), and
reads:
\begin{eqnarray}
\label{self_energy_out2}
\Sigma_{1,out}^{(\infty)} = { e^2 \over 4 \epsilon_m x_0}
\int_{R/x_0}^{1}{d\xi_\bot} \xi_\bot \left[ K_0^2(\xi) + K_1^2(\xi) \right].
\end{eqnarray}
In the lowest order in $R/x_0$, the result is:
\begin{eqnarray}
\label{self_energy_outout}
\Sigma_{1,out}^{(\infty)} = \Sigma_{a,in}^{(\infty)} - 0.13 {e^2 \over 2 \epsilon_m x_0},
\end{eqnarray}
where the second term is a boundary correction, due to the cylindrical
geometry. Including this boundary correction, the image self-energy, in the lowest order in $R/x_0$, reads:
\begin{eqnarray}
\label{self_energy_im}
\Sigma_{1}^{(\infty)} = {1 \over 2}{e^2 \over \epsilon_m x_0} \left( \log(2x_0/R) - \gamma \right) - 0.13 {e^2 \over 2 \epsilon_m x_0},
\end{eqnarray}
where the first term returns us back to Eq.~(\ref{self_energy_along_pore}).

We now proceed to determine the part of the self-energy,
$\Sigma_{3}^{(\infty)}$, at distances $x \gg x_0$. This reads:
\begin{eqnarray}
\label{self_energy_ext}
\Sigma_{3}^{(\infty)} = { \epsilon_m
\over 8 \pi} \int_{ext}{d{\mathcal V}} ({\bf E^{ext}})^2,
\end{eqnarray}
where the external volume is beyond the cylinder of length $2x_0$
and radius $x_0$. With the help of Eq.~(\ref{electric_ext}), this
reads:
\begin{eqnarray}
\label{self_energy_ext2}
\Sigma_{3}^{(\infty)} = {\pi \over 8} {e^2  \over 2 \epsilon_m x_0},
\end{eqnarray}
and should be contrasted with the naive expression:
$\Sigma_{3}^{(\infty)} = {e^2 / 2 \epsilon_m x_0}$, based on a
spherical symmetry.

From Eqs.~(\ref{self_energy_im}) and (\ref{self_energy_ext2}), the
total image self-energy, Eq.~(\ref{self_energy_inout}), to the
lowest order in $R/x_0$, reads:
\begin{eqnarray}
\label{self_energy_inout_final}
\Sigma^{(\infty)} = {1 \over 2}
{e^2 \over \epsilon_m x_0} \left( \log(2x_0/R) - \gamma \right) +
0.26{ e^2  \over 2 \epsilon_m x_0}.
\end{eqnarray}
%


\section{Numerical estimates for the infinite channel}
\label{estimates}

The most straightforward and accurate way, of determining the
value of the self-energy of an ion, Eq.~(\ref{self_energy_def}),
has been taken by Parsegian who has used a formula~\cite{Sm},
similar to Eqs.~(\ref{phi_real}) and (\ref{phi_fourier}), which
has been substituted in Eq.~(\ref{self_energy_def}), and evaluated
numerically. With the help of our Eqs.~(\ref{phi_real}) and
(\ref{phi_fourier}), Eq.~(\ref{self_energy_def}) reads:
\begin{eqnarray}
\label{SE_parsegian}
\Sigma^{(\infty)} = \int_{0}^{+\infty}{{dk \over \pi}}
{e^2 \over \epsilon_w} {(\epsilon_w - \epsilon_m)K_0(kR)K_1(kR) \over \epsilon_w K_0(kR)I_1(kR) +
\epsilon_m K_1(kR)I_0(kR)},
\end{eqnarray}
where the energy is measured with respect to the Born energy, and
may be evaluated with any standard software. We shall refer to
this solution, for the image self-energy, as the {\it exact
numerical solution}.

Our analysis of the previous sections, on the other hand, has
revealed some important features of the electrostatics of ion
channels, in particular the crossover length $x_0$, the 1D nature of the Coulomb potential
and expressions for all quantities of interest,
including the self-energy of Eq.~(\ref{SE_parsegian}) and the main
contribution to this image self-energy, the 1D Coulomb part.
We proceed to evaluate these expressions, by taking the case of,
{\it e.g.} the model-pore gramicidin A which has a radius $R=3.0
{\mathrm{\AA}}$ and a length $L=25.0 {\mathrm{\AA}}$, and for the
typical room-temperature values: $\epsilon_m = 2$ and $\epsilon_w
=80$.

The crossover length, $x_0$, has to be determined
self-consistently from Eq.~(\ref{x_0}). In the logarithmic
approximation, this length reads:
\begin{equation}
\label{x_0_3}
x_0 \approx R \sqrt{{\epsilon_w \over 2 \epsilon_m}}\sqrt{0.5\log(2\epsilon_w /\epsilon_m) - \gamma}.
\end{equation}
Using a more precise graphical method, cf. Fig.~\ref{lengthx02}, yields:
\begin{equation}
\label{x_0_value}
x_0 \approx 6.3 R,
\end{equation}
where we have used the fact that $\epsilon_m = 2$ and $\epsilon_w
=80$. In the case of the model-pore gramicidin A ($R=3.0
{\mathrm{\AA}}$): $x_0 \approx 19.0 {\mathrm{\AA}}$. The peculiar
potential of Eq.~(\ref{phi_real_series_smallx}) plays a dominant
role over this length, from the origin. Beyond the length $x_0$,
the 1D Coulomb potential crosses over to the 3D one. We have therefore to compare the length of the channel,
$L=25.0 {\mathrm{\AA}}$, to the region where the anti-confinement
dominates: $2x_0 \approx 38.0 {\mathrm{\AA}}$.

Obviously, such a value of $2x_0 > L$ shows that the
1D Coulomb potential dominates in the model nano-pore. We
shall come back on this crucial fact, in the next section, when
dealing with finite-size effects. Substituting the expression of
$x_0$, in Eq.~(\ref{phi_real_series_smallx2}), leads to:
\begin{eqnarray}
\label{conf_potential}
&&\varphi(x) \approx
{e \over \sqrt{ \epsilon_w \epsilon_m} R} \sqrt{ 2\left( \log(2x_0/R) - \gamma \right)}
\exp(- {|x| / x_0}),
\nonumber \\
&&R \ll x \ll x_0,
\end{eqnarray}
which displays the "geometric mean" screening: $\propto \sqrt{
\epsilon_w \epsilon_m}$, and is affected by a logarithmic factor
characteristic of the intrinsic 1D nature of the system (see also
Ref.~\onlinecite{PF} where the geometric-like screening has been considered qualitatively).

\begin{table}
\caption{\label{Table} Image self-energy, of an ion in an infinite
channel, in various units from the exact numerical calculation,
see Eqs.~(\ref{phi_real}), (\ref{phi_fourier}) and discussion
below them as well as Ref.~\onlinecite{Parsegian}, and from our
asymptotic formulas, Eq.~(\ref{self_energy_inout_final}). We have
taken $\epsilon_m = 2$ and $\epsilon_w =80$.}
\begin{ruledtabular}
\begin{tabular}{lllll}
Self-energy unit&                       Exact&      Eq.~(\ref{self_energy_inout_final})&  \\
\hline
$e^2/\sqrt{\epsilon_w \epsilon_m}R$&    $1.08$&         $1.11$&   \\
$e^2/\epsilon_m R$&                     $0.17$&         $0.17$&   \\
$e^2/\epsilon_w R$&                     $6.80$&         $7.01$&   \\
${k_BT}$&              $16.0$&     $16.6$&     ($R=3.0 {\mathrm{\AA}}$)      \\
\end{tabular}
\end{ruledtabular}
\end{table}
As a consequence of Eq.~(\ref{conf_potential}), if an ion is in
the pore, any additional ion will feel a constant repulsive force,
cf. Eq.~(\ref{conf_force}).
Substituting the expression of $x_0$, cf. Eq.~(\ref{x_0}), in
Eq.~(\ref{conf_force}), this force simply reduces to:
\begin{eqnarray}
\label{conf_force2}
F = {2 e^2 \over \epsilon_w R^2}.
\end{eqnarray}
This is an unrealistically large (because we consider an infinite channel)
force: $F_{a} \approx 48.9 ~{\mathrm{eV}} / {\mathrm{\AA}}$.
The strong repulsion originating from
Eq.~(\ref{conf_force2}) is therefore responsible for the large
image self-energy of the charge, cf.
Eq.~(\ref{self_energy_along_pore}), which, with the help of
Eq.~(\ref{x_0}), and to the lowest order in $R/x_0$, reads:
\begin{eqnarray}
\label{self_energy_total_electric_field}
\Sigma_{1}^{(\infty)} = {1 \over 2} {e^2 \over \sqrt{\epsilon_w \epsilon_m} R} \sqrt{2( \log(2x_0/R) - \gamma)} - C_g,
\end{eqnarray}
where $C_g = 0.13 (e^2 / 2 \epsilon_m x_0)$, is the geometric
correction, appearing in Eq.~(\ref{self_energy_im}).  With the
help of $\epsilon_m = 2$ and $\epsilon_w =80$,
Eq.~(\ref{self_energy_total_electric_field}) yields:
\begin{eqnarray}
\label{ISE_value}
\Sigma_{1}^{(\infty)} \approx (44.02/R - 2.90/R) {k_BT}
\approx (41.12/R) {k_BT},
\end{eqnarray}
where $C_g \approx (2.90/R) {k_BT}$. This correction is
not negligible when it comes to compare our results with precise
numerical estimates. For $R=3 {\mathrm{\AA}}$, $C_g \approx 0.97
k_B {T}$, and this yields: $\Sigma_{1}^{(\infty)}
\approx 13.7 k_B {T}$. For $R=2 {\mathrm{\AA}}$, a
drastic increase results: $\Sigma_{1}^{(\infty)} \approx 20.56
k_B {T}$, where $C_g \approx 1.45 k_B {T}$. On
the other hand, for $R=5 {\mathrm{\AA}}$, this quantity is reduced
to: $\Sigma_{1}^{(\infty)} \approx 8.22 k_B {T}$, $C_g
\approx 0.58 k_B {T}$.

To estimate the total image self-energy one should consider
Eq.~(\ref{self_energy_inout_final}), which takes into account of
the large-distance part of the energy. Adding this contribution,
to Eq.~(\ref{ISE_value}), yields:
\begin{eqnarray}
\label{SE_value} \Sigma^{(\infty)} \approx (49.82/R) k_B T.
\end{eqnarray}
For $R = 3 {\mathrm{\AA}}$, Eq.~(\ref{SE_value}), yields:
$\Sigma^{(\infty)} \approx 16.6 k_B {T}$.
Numerical estimates for $\Sigma^{(\infty)}$, in various units, are
displayed in Tab.~\ref{Table}. They agree with the
exact results of numerical simulations.

Notice that, to express the self-energy in units of $k_B
{T}$, where $k_B$ is Boltzmann's constant, we have found
it convenient to introduce Bjerrum's length, $l_B$. At room
temperature and in water, Bjerrum's length reads: $l_B = e^2/
\epsilon_w k_B {T}$. To be able to compare safely our
results with previous studies~\cite{Levitt,Jordan} we take the
value $l_B \approx 7.038 {\mathrm{\AA}}$ (Jordan converts the
values of Levitt (in $k_B {T}$) in units of
$e/\epsilon_wR$ and this value of $l_B$ agrees with the
conversion).


\section{Finite-size effects: translocation energy
in nano-channels}
\label{finite_sizes}

In the case of a finite-length channel, calculations similar to
the infinite length case lead to coupled non-linear integral
equations, at the level of the boundary conditions imposed on the
electrostatic potential of the inhomogeneous Poisson equation.
Even an approximate solution of these boundary equations seems
intractable, analytically, and numerical simulations appear to be
necessary. Such a numerical task has been undertaken in Refs.
\onlinecite{Levitt} and~\onlinecite{Jordan}, by two different
numerical methods, showing that the self-energy
$\Sigma^{(\infty)}$ becomes of the order of $\Sigma^{(L)} \approx
6.7 k_B {T}$ for a length $L=25.0 {\mathrm{\AA}}$ and a
radius $R = 3 {\mathrm{\AA}}$. The latter is defined as:
\begin{equation}
\label{Self_L_def}
\Sigma^{(L)} = (e/2)(\varphi(0)- \varphi(L/2)),
\end{equation}
which is simply the work required to bring the charge from the
entrance of the channel, at $x=L/2$, to its center, at $x=0$.
Two formulas were then proposed to take into account of the
finite size effects.
\begin{figure}
\includegraphics[width=8cm,height=4cm]{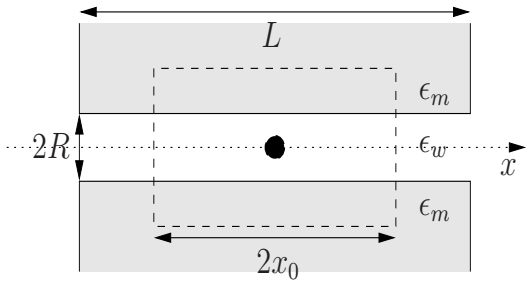}
\caption{ \label{LC} Schematic view of a long channel, for which: $L \gg 2x_0$, cf. Eq.~(\ref{x_0}).}
\end{figure}
\begin{figure}
\includegraphics[width=8cm,height=4cm]{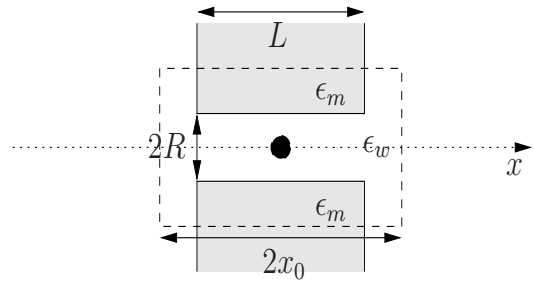}
\caption{ \label{SC} Schematic view of a nano-length channel, for which: $L \ll 2x_0$, cf. Eq.~(\ref{x_0}). This case corresponds to the typical channels we are interested in, {\it e.g.} gramicidin A for which $L \approx 2.5 {\mathrm{nm}}$, $R \approx 3 {\mathrm{\AA}}$ and $2x_0 \approx 12.6 R \approx 3.8 {\mathrm{nm}}$.}
\end{figure}

For large lengths Parsegian has proposed~\cite{Levitt}:
\begin{eqnarray}
\label{self_energy_finite1a}
\Sigma^{(L)} \approx \Sigma^{(\infty)} - {e^2 \over \epsilon_m L} \log({2 \epsilon_w \over \epsilon_m + \epsilon_w}),
\end{eqnarray}
where the first term is the self-energy of the infinite channel
and the second term corresponds to the self-energy of a charge in
a membrane without a pore. On the light of the previous sections,
this "large length" ansatz, should work for lengths larger than $2
x_0$. We may understand this result on the basis of
Fig.~\ref{LC}, by decomposing the self-energy into two parts,
while bringing the charge from infinity to the origin:
\begin{eqnarray}
\label{self_energy_finite1aa}
\Sigma^{(L)} \approx \Sigma_{a}^{(\infty)} + {e^2 \over 2 \epsilon_m}({1 \over x_0} - {2 \over L}),
\end{eqnarray}
where the first term corresponds to the 1D Coulomb part of
the self-energy of the infinite channel and the second, to the
energy required to displace the charge from $L/2$ to $x_0$ (for the finite channel image charges do not play any role beyond $L/2$ from the origin and the corresponding self-energy contribution has been neglected). Up to numerical coefficients, the first two
terms yield $\Sigma^{(\infty)}$ and the last term yields the
(reducing) effect of the membrane. Unfortunately, Eq.~(\ref{self_energy_finite1a}) seems to work only for
unrealistically long channels, well beyond $2x_0$, {\it i.e.} the percentage of error, with respect to numerical simulations, becomes less than $5\%$ for lengths larger than $40 R$.

For the most relevant case of small length channels, cf. Fig.~\ref{SC},  Jordan has proposed, without any justification other than curve-fitting, the following expression~\cite{Jordan}:
\begin{eqnarray}
\label{self_energy_finite1b}
\Sigma^{(L)} \approx \Sigma^{(\infty)}[1-\exp(-b \delta)],
\end{eqnarray}
where $b=0.131$ and $\delta$ is the reduced channel length,
$\delta = L/2R$ . We see immediately that the characteristic length
appearing in the exponential corresponds to: $L_0 \approx 2.44
x_0$. The "small-length" case will therefore obviously correspond
to the regime where $L \ll 2 x_0$. We now proceed to justify the
curve-fitting Eq.~(\ref{self_energy_finite1b}) and improve on the
result of Jordan.

As is clear from the previous lines, the knowledge of $x_0$ allows
us to consider that channels of characteristic length: $L \ll 2
x_0$, are finite, with respect to the 1D Coulomb part
of the self-energy, whereas channels of characteristic lengths: $L
\gg 2 x_0$, are formally infinite with respect to the same
quantity. As a matter of fact, the 1D Coulomb part of the self-energy of a channel of
length $L \gg 2 x_0$ saturates: $\Sigma_{1}^{(L)} \approx
\Sigma_{a}^{(\infty)}$. On the other hand, this self-energy, for
a channel of length $L \ll 2 x_0$, is determined with the help of
Eq.~(\ref{conf_potential}) and Eq.~(\ref{Self_L_def}). The first term
in Eq.~(\ref{Self_L_def}) corresponds to $\Sigma_{1}^{(\infty)}$ and the second gives the
leading correction in $L/x_0$. Actually, higher order
corrections in $L/2 x_0$ are important. As in
Eq.~(\ref{conf_potential}), such corrections may be
re-summed in an exponential factor. This yields:
\begin{eqnarray}
\label{self_energy_finiteI2b}
\Sigma_{1}^{(L)} \approx \Sigma_{1}^{(\infty)} [ 1 - \exp(-L/2x_0)], \quad L \ll 2 x_0,
\end{eqnarray}
where $\Sigma_{1}^{(\infty)}$ is given by Eq.~(\ref{self_energy_total_electric_field}), which justifies the functional form of Jordan's expression on a
physical basis. In the lowest order in $L/2 x_0$,
$\Sigma_{1}^{(L)}$, in Eq.~(\ref{self_energy_finiteI2b}),
increases linearly with $L$, a feature of the 1D nature of the Coulomb potential:
$\Sigma_{1}^{(L)} \approx \Sigma_{1}^{(\infty)} L / 2 x_0$.
For the case of interest ($\epsilon_m = 2$ and $\epsilon_w =80$),
the image self-energy of a channel of length $L$ and radius $R$
reads:
\begin{eqnarray}
\label{translocation}
&&\Sigma_{1}^{(L)}(L,R) \approx  ({44.02 \over R} - C_g)
[ 1 - \exp[-\delta/6.3]]~k_B {T}, \nonumber \\
&&L \ll 2 x_0,~~~~
\end{eqnarray}
where $\delta = L/2R$ and $C_g \approx (2.90/R) k_B {T}$ from
the previous section.
Actually, Eq.~(\ref{translocation}) works better than
Eq.~(\ref{self_energy_finite1a}) at lengths $L \approx 2x_0$. This implies
that, in Eq.~(\ref{self_energy_finite1a}), corrections due to the
presence of the membrane beyond $|x| > x_0$ manifest only for very large lengths, $L \gg 2 x_0$,
which is beyond the lengths of the channels we are interested in.

We also notice from Eq.~(\ref{translocation}), that the
self-energy cannot be expressed in terms of a single scaling
parameter, here $\delta = L/2R$. Eq.~(\ref{translocation}) depends
separately on $L$ and $R$. This fact is very clear from the work
of Levitt, that we have reproduced in Tab.~\ref{TableFinite}. For
a given parameter $\delta$, {\it e.g.} $\delta = 6.25$, the
channel $(L=25{\mathrm{\AA}},~R=2{\mathrm{\AA}})$ is characterized
by $\Sigma^{(L)} \approx 13.3 k_B {T}$ whereas the channel
$(L=37.5 {\mathrm{\AA}},~R=3 {\mathrm{\AA}})$ is
characterized by: $\Sigma^{(L)} \approx 8.91 k_B {T}$.
This is a drastic energy variation, which is well reproduced by
Eq.~(\ref{translocation}).

Eq.~(\ref{translocation}) gives therefore a fit-parameter free
expression for the self-energy of a charge in nano-channels, in the range of validity: $R \ll L \ll 2x_0$. As can be seen from Tab.~\ref{TableFinite}, the average percentage of error (in the regime of validity) of Eq.~(\ref{translocation}), with respect to numerical simulations, is: $2.16\%$ for the data of column DL and PJ1 and of $4.14\%$ for column PJ2. With respect to the data of Levitt~\cite{Levitt} we are within $3.\%$ of error and with respect to the data of Jordan~\cite{Jordan} within $5.\%$, {\it i.e.} similar than the error bar of Jordan's
curve-fitting formula. Notice also that the accuracy extends
beyond $L=2x_0$ (or $\delta=\delta_0 = 6.3$), with an error still within $3.\%$, with respect to PJ1 and within $6.\%$, with respect to PJ2. In the
limit $L \gg 2 x_0$, Eq.~(\ref{translocation}) also gives a better accuracy than Eq.~(\ref{self_energy_finite1a}) the values of which have been displayed in the last column of Tab.~\ref{TableFinite}. Better accuracy, with respect to the numerical simulations, may be reached if $C_g$ is considered as a fitting parameter but we shall not dwell on this point as our fit-free results have already a very good accuracy.
\begin{table*}
\caption{\label{TableFinite} Image self-energy of an ion in
a channel of length $L$, in units of $k_B {T}$, as a
function of the reduced channel length $\delta = L/2R$, where $R$
is the radius of the channel. We have taken $\epsilon_m = 2$ and
$\epsilon_w =80$. Energies are in units of $k_B
{T}$ and lengths are in units of ${\mathrm{\AA}}$. The
crossover length, $L_0=2x_0$ between large and small channels,
corresponds to the universal value (independent of the length and
radius of the channel): $\delta_0 = 6.3$. Our asymptotic
calculations are valid for lengths: $R < L < 2x_0$.
Eq.~(\ref{translocation}), is taken with no fitting parameter. The percentage of error appearing in columns DL, PJ1 and PJ2, is that of Eq.~(\ref{translocation}) with respect to the numerical simulations.}
\begin{ruledtabular}
\begin{tabular}{llllllllllll}
L &      R &      $\delta$&       DL~\footnotemark[1]&          PJ1~\footnotemark[2]&       PJ2~\footnotemark[3]&       Eq.~(\ref{translocation})&  Eq.~(\ref{self_energy_finite1a})&   \\
\hline
9&              3&              1.5&        -&  $2.58~(11.0\%)$&        $2.51~(13.4\%)$&        $2.90$&     -&\\
15&             3&              2.5&        $4.34~(3.3\%)$&     -&      $4.28~(4.7\%)$&     $4.49$&     $3.45$&\\
21&             3&              3.5&        -&      $5.86~(0.3\%)$&     $5.82~(0.3\%)$&     $5.84$&     $7.04$&\\
25&     3&      4.17    &       $6.72~(1.4\%)$&     -&              -&              $6.63$&     $8.47$&\\
30&             3&               5.&        -&      $7.69~(2.4\%)$&     $7.72~(2.8\%)$&     $7.51$&     $9.73$&\\
35&     3&          5.83&           $8.48~(2.5\%)$&     -&              -&              $8.27$&     $10.62$&\\
37.5&           3&              6.25&       -&      $8.91~(3.4\%)$&     $9.00~(4.4\%)$&     $8.62$&     $10.98$&\\
50&     3&      8.33&       $10.25~(2.0\%)$&    -&      $10.59~(5.4\%)$&    $10.05$&    $12.24$&\\
52.5&       3&      8.75&       -&      $10.45~(1.6\%)$&    $10.87~(5.6\%)$&    $10.29$&    $12.42$&\\
75&     3&      12.5&       -&      $11.53~(2.5\%)$&    $12.50~(5.8\%)$&    $11.82$&       $13.49$& \\
$\infty$&       3&              $\infty$&       $16$&           -&              -&              $14.70$&    $16$&   \\
\\
15&     2&      3.75    &       $9.3~(0.1\%)$&          -&                 -&            $9.22$&   \\
25&     2&      6.25    &       $13.3~(2.8\%)$&         -&                 -&            $12.94$&   \\
35&             2&              8.75    &       $15.6~(0.1\%)$&         -&                 -&            $15.43$&   \\
50&             2&          12.5    &       $17.2~(3.0\%)$&         -&                 -&            $17.73$&   \\
$\infty$&       2&              $\infty$&       $24$&           -&                 -&            $20.56$&   \\
\\
15&     5&      1.5 &   $1.54~(11.5\%)$&    -&        -&            $1.74$&   \\
25&     5&      2.5 &   $2.59~(3.7\%)$&     -&        -&            $2.69$&   \\
35&     5&      3.5 &   $3.50~(0.3\%)$&     -&        -&            $3.51$&   \\
50&     5&      5   &   $4.59~(1.8\%)$&     -&                -&            $4.51$&   \\
$\infty$&       5&              $\infty$&       $9.62$&         -&                -&            $8.22$&   \\

\footnotetext[1]{As determined from Table I of
Levitt~\cite{Levitt}.}

\footnotetext[2]{As determined from Table I, column 2, of Jordan~\cite{Jordan}. }

\footnotetext[3]{As determined from Table I, column 3, of Jordan~\cite{Jordan}.}

\end{tabular}
\end{ruledtabular}
\end{table*}
%


\section{Conclusion}
\label{conclusion}

In summary, we have shown that finite-$T$ transport, across ionic
channels, is determined by zero-temperature (as there is no
screening~\cite{KZLS} by counter-ions within the channels we are interested in) classical electrostatics. Our main results have shown that ions in the water channel of a lipid membrane interact via a one-dimensional Coulomb potential, cf. Eq.~(\ref{phi_real_series_smallx}), below a crossover length $x_0$, cf. Eq.~(\ref{x_0}), and a three-dimensional Coulomb potential beyond this length. The length $x_0$ has
been shown to be of the order of the length of, {\it e.g.} the model pore gramicidin A. An expression for the image self-energy of a finite nano-channel has then been obtained, cf. Eqs.~(\ref{self_energy_finiteI2b}) and (\ref{self_energy_total_electric_field}), with no fitting parameter and an agreement within an error of $5\%$ with the existing numerical simulations available from the literature on the subject.
Combining Eqs.~(\ref{self_energy_finiteI2b}) and (\ref{self_energy_total_electric_field}) the most general expression for the image self-energy of a channel of length $L$ and radius $R$ reads:
\begin{eqnarray}
\label{translocation_energy}
\Sigma_{1}^{(L)}(L,R) = \qquad ~~~~~~~~ \qquad \nonumber \\
{\mathcal{N} \over 2} {e^2 \over \sqrt{\epsilon_w \epsilon_m} R} \sqrt{2( \log(2x_0/R) - \gamma)}[1 - \exp(-L/2x_0)], \nonumber \\
R \ll L \ll 2x_0, \qquad ~~~~~~~~ \qquad
\end{eqnarray}
where $\gamma \approx 0.577$ is Euler's constant, $x_0$ is determined by Eq.~(\ref{x_0}) and $\mathcal{N}=1-0.13$, is a numerical coefficient originating from $C_g$, cf. Eq.~(\ref{self_energy_total_electric_field}). The latter may be of the order of a few $k_BT$ (depending on the characteristics of the channel under consideration) and has to be taken into account for numerical estimates. In the lowest order in $L/2x_0$,
Eq.~(\ref{translocation_energy}) increases linearly with the length of
the channel, a feature of the 1D Coulomb potential.

The image self-energy of Eq.~(\ref{translocation_energy}) corresponds to the {\it translocation}, or activation, energy through a finite length channel, {\it e.g.} the energy required to displace a charge from the entrance of a channel to its center. In relation with experiments, the existence of such electrostatic barrier requires a threshold voltage difference,
$e \Delta V_c(L,R) =  \Sigma_{1}^{(L)}(L,R)$ (where the self-energy is in units of $eV$), across the channel in order to pass an ion through it (no chemical potential difference is taken into account here). Equivalently, this corresponds to the existence of a threshold electric field: $E_c = \Delta V_c(L,R) / L$, applied along the length of the channel. As an inverse problem, measurement of this threshold field would then give access to information on the channel, its radius or length, with the help of
Eq.~(\ref{translocation_energy}).

The image self-energy of Eq.~(\ref{translocation_energy}) determines the main exponential dependence of the conductance of a nano-scale channel:
\begin{eqnarray}
\label{conductance}
G \propto \exp[- \Sigma_{1}^{(L)}(L,R)/ k_B T].
\end{eqnarray}
Up to now, all our estimations where performed at room temperatures
for $\epsilon_w = 80$. Actually, the dielectric constant of water is sensitive to temperature and varies from $\epsilon_w(273~K) = 87.78$ to
$\epsilon_w(323~K) = 69.91$. Basic estimations of $\Sigma_{1} \equiv \Sigma_{1}(T)$ which take into account of this temperature variation of $\epsilon_w$ show that the ratio $\Sigma_{1}(T) / k_B T$ becomes approximately temperature - independent (the coefficient $C_T$ defined as: $\Sigma_{1}(T) / T|_{T=273K} = C_T \Sigma_{1}(T) / T|_{T=323K}$ is
equal to $C_T \approx 1.18$ for a temperature-independent dielectric constant and to $C_T \approx 1.00$ with the above temperature-dependence of $\epsilon_w$). This implies that, despite the fact that the conductance is affected by an exponentially small factor, its thermal dependence is weak so that the ionic flow through channels of cell membranes is stable, being approximately constant over more than two decades of units of temperature, around room temperature.

\acknowledgments

I am grateful to B.~I.~Shklovskii for fruitful discussions which initiated this work. I thank A.~V.~Finkelstein, D.~N.~Ivankov and A.~M.~Dykhne for sharing their version of the expression of the image self-energy for the infinite channel case, cf. Ref.~\onlinecite{FID}. I also thank G.~Huber for general comments on the manuscript. This work is supported by INTAS Grant No.~2212.


\end{document}